\documentclass[prl,noshowpacs,english,superscriptaddress,twocolumn]{revtex4}
\usepackage{amsmath,amssymb,amsfonts,mathrsfs}
\usepackage{amsthm}
\usepackage{CJK}
\usepackage{bm}
\usepackage{babel}
\usepackage{graphicx}
\allowdisplaybreaks
\usepackage{braket}
\usepackage[breaklinks]{hyperref}
\usepackage{color}
\hypersetup{colorlinks=true, linkcolor=blue, citecolor=blue, filecolor=blue, urlcolor=blue}
\usepackage{booktabs}

\begin{document}

\title{Intrinsic second-order topological insulators in two-dimensional polymorphic graphyne with sublattice approximation}

\author{Z. J. Chen}
\affiliation{Songshan Lake Materials Laboratory, Dongguan 523000, People's Republic of China}
\affiliation{Beijing National Laboratory for Condensed Matter Physics and Institute of Physics, Chinese Academy of Sciences, Beijing 100190, People's Republic of China}
\author{S. G. Xu}
\affiliation{Department of Physics $\&$ Institute for Quantum Science and Engineering, Southern University of Science and Technology, Shenzhen 518055, People's Republic of China}
\affiliation{Quantum Science Center of Guangdong-Hong Kong-Macao Greater Bay Area (Guangdong), Shenzhen 518045, People’s Republic of China}
\author{Z. J. Xie}
\affiliation{International School of Microelectronics, Dongguan University of Technology, Dongguan 523808, People's Republic of China}
\author{H. Xu}
\email[]{xuh@sustech.edu.cn}
\affiliation{Department of Physics $\&$ Institute for Quantum Science and Engineering, Southern University of Science and Technology, Shenzhen 518055, People's Republic of China}
\affiliation{Quantum Science Center of Guangdong-Hong Kong-Macao Greater Bay Area (Guangdong), Shenzhen 518045, People’s Republic of China}
\author{H. M. Weng}
\email[]{hmweng@iphy.ac.cn}
\affiliation{Songshan Lake Materials Laboratory, Dongguan 523000, People's Republic of China}
\affiliation{Beijing National Laboratory for Condensed Matter Physics and Institute of Physics, Chinese Academy of Sciences, Beijing 100190, People's Republic of China}

\begin{abstract}
In two dimensions, intrinsic second-order topological insulators (SOTIs) are characterized by topological corner states that emerge at the intersections of distinct edges with reversed mass signs, enforced by spatial symmetries. Here, we present a comprehensive investigation within the class BDI to clarify the symmetry conditions ensuring the presence of intrinsic SOTIs in two dimensions. We reveal that the (anti-)commutation relationship between spatial symmetries and chiral symmetry is a reliable indicator of intrinsic corner states. Through first-principles calculations, we identify several ideal candidates within carbon-based polymorphic graphyne structures for realizing intrinsic SOTIs under sublattice approximation. Furthermore, we show that the corner states in these materials persist even in the absence of sublattice approximation. Our findings not only deepen the understanding of higher-order topological phases but also open new pathways for realizing topological corner states that are readily observable.
\end{abstract}


\maketitle

The exploration of topological insulators (TIs) has significantly advanced our understanding of condensed matter physics over recent decades \cite{hasan2010colloquium,qi2011topological,kane2005quantum,bernevig2006quantum,zhang2009topological}, while current research on higher-order topological insulators (HOTIs) marks a new frontier in the study of topological phases \cite{benalcazar2017quantized,benalcazar2017electric,schindler2018higher,song2017d,langbehn2017reflection}. HOTIs are characterized by unique in-gap states localized at higher co-dimensional boundaries, namely corner or hinge states \cite{schindler2018higher}. These states are termed ``intrinsic" when they arise from spatial symmetries that reverse the mass term at edges or surface intersections \cite{queiroz2019partial,geier2018second}. This reversal, governed by bulk symmetries, ensures the presence of corner or hinge states as long as the bulk band gap remains open \cite{geier2018second}. In contrast, extrinsic HOTIs lack this bulk protection, so that the presence of their corner or hinge states depend on termination selection \cite{geier2018second}.

In two-dimensional (2D) systems, higher-order topological states are uniquely confined to corners, defining what are known as second-order topological insulators (SOTIs) \cite{schindler2018higher}. It has been shown that the topological classification of intrinsic corner states aligns with certain Altland-Zirnbauer classes in one dimension \cite{langbehn2017reflection}. Consequently, intrinsic corner states can only exist within five out of the ten Altland-Zirnbauer classes, $i.e.$, AIII, BDI, D, DIII, and CII, where either chiral or particle-hole symmetry is essential  \cite{langbehn2017reflection,chiu2016classification}. From another perspective, corner states may shift towards bulk states unless chiral or particlehole symmetry anchor them precisely at the zero energy \cite{hwang2019fragile,khalaf2021boundary}. This symmetry requirement poses significant challenges in realizing intrinsic second-order topological states in 2D materials. Notably, recent studies have reported the emergence of SOTIs in two dimensions, thereby expanding the understanding of higher-order topological phases \cite{liu2019two,chen2021graphyne,sheng2019two,lee2020two,mu2022kekule,qian2022c,arroyo2023fractional,ma2023obstructed,nunez2023higher,qian2021second,xue2021higher,hitomi2021multiorbital,li2022second,hu2022intrinsic,guo2022quadrupole,pan2022two,zhao2021higher,chen2020universal,liu2022second,bai2023engineering,cai2023second,mao2023ferroelectric,li2023realization}. However, within these materials, chiral and particle-hole symmetries, emerging as approximations, have been observed in a few cases, notably in graphyne \cite{liu2019two,chen2021graphyne} and graphdiyne \cite{sheng2019two,lee2020two,mu2022kekule}. In other scenarios, such as fragile topological insulators (FTIs) \cite{ahn2018band,wang2019higher,benalcazar2019quantization,bouhon2019wilson,bouhon2020geometric}, obstructed atomic insulators (OAIs) \cite{kruthoff2017topological,bradlyn2017topological,gao2022unconventional,xu2021three,xu2021filling}, and boundary-obstructed topological phases (BOTPs) \cite{khalaf2021boundary,ezawa2020edge}, this symmetry requirement  is not necessary, and corner states may hybridize with bulk states or disappear as gaps close within valence bands or edge states \cite{khalaf2021boundary}. 

In this work, we contribute to this dynamic field by examining the symmetry conditions necessary for intrinsic 2D SOTIs, focusing specifically on class BDI, and exploring their material realization based on sublattice approximation. By analyzing the transformation of an angle-dependent edge Hamiltonian under 2D point group symmetries, we identify potential positions for intrinsic corner states, which depend on whether the spatial symmetries commute or anticommute with the chiral symmetry. We also identify several ideal candidates in polymorphic graphyne for realizing intrinsic 2D SOTIs with sublattice approximation, highlighting a promising strategy for achieving observable topological corner states. The signatures of higher-order topological phases are revealed through first-principles calculations, offering potential for observing corner states in 2D SOTIs.

\begin{figure}[!t]
	\centering
	\includegraphics[scale=0.068]{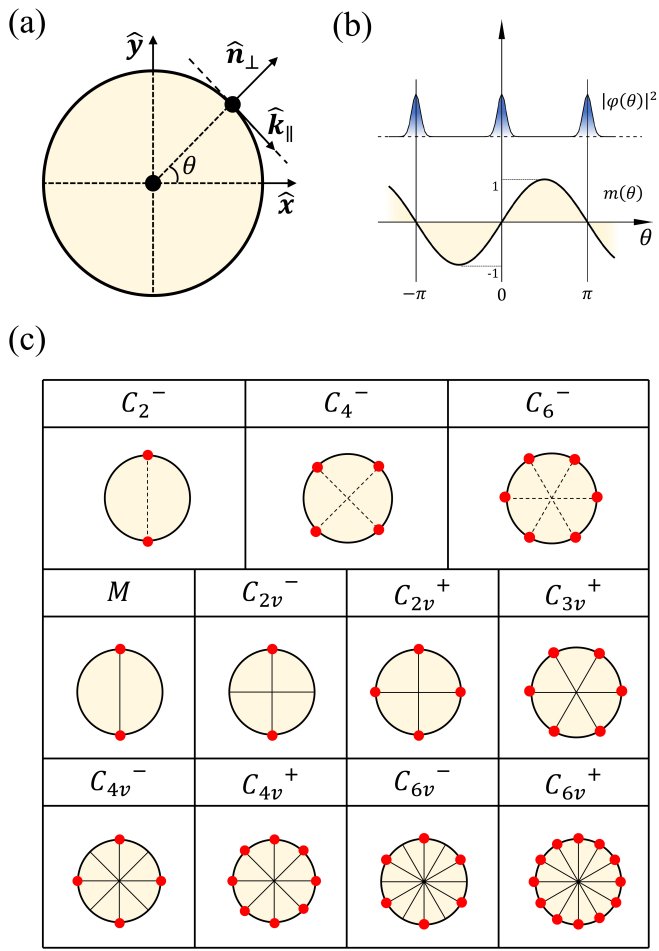}
    \caption{(a) Diagram of a two-dimensional lattice with an edge that varies with $\theta$, defined as the angle between the normal direction of the edge and the $\hat{x}$-direction. (b) Graph showing the variation of the
    probability density of the edge wave function and the mass parameter as a function of $\theta$. (c) Diagram depicting the potential positions of intrinsic corner states in two-dimensional point groups within class BDI. A superscript $+(-)$ indicates that minimal rotation symmetry commutes (anticommutes) with chiral symmetry. Solid lines represent mirror planes, while dashed lines serve as reference lines.}
\label{fig-1}
\end{figure}

Class BDI is a category in the ten-fold way classification scheme defined by three non-spatial symmetries: time-reversal ($\mathcal{T}$), particle-hole ($\mathcal{P}$), and chiral ($\mathcal{C}$) symmetries, with each satisfying the conditions $\mathcal{T}^2=1$, $\mathcal{P}^2=1$, and $\mathcal{C}^2=1$ \cite{langbehn2017reflection,chiu2016classification}.  To clarify the symmetry conditions of intrinsic 2D SOTIs in class BDI, we begin with a 2D lattice designed to allow cuts in arbitrary directions, as shown in Fig. \ref{fig-1} (a). While spatial symmetries are broken at the boundary of this lattice, the three non-spatial symmetries retain their constraints. This results in edge states that, depending on the cut angle $\theta$, mimic one-dimensional lattices categorized under class BDI. We employ a minimal Dirac Hamiltonian with only one mass term to describe the boundary states, represented as: 
\begin{equation}
H_\theta(k_\parallel) = m(\theta)\sigma_x + k_\parallel\sigma_y,
\end{equation}
where $\sigma_i (i=x, y, z)$ are Pauli matrices, $k_\parallel$ is the momentum parallel to the edge, and $m$($\theta$) is a mass parameter. The Hamiltonian respects time-reversal symmetry $\mathcal{T}=K$ and anticommutes with chiral symmetry $\mathcal{C}=\sigma_z$. For spatial symmetries changing $\theta$, the Hamiltonian undergoes transitions described by:
\begin{equation}
U_R^\dagger H_\theta(k_\parallel)U_R = H_{R\theta}(\eta k_\parallel)
\end{equation}
where $R$ represents a spatial symmetry, $U_R$ is its matrix representation, and $\eta$ takes the values $\pm1$ to indicate whether $R$ is a proper or improper rotation. Akin to first-order topological boundary states that exist as domain walls between the bulk and vacuum, intrinsic corner states appear as domain walls along the edge at positions where the mass term changes sign, as shown in Fig. \ref{fig-1} (b). The presence of these states depends on finding a matrix representation of spatial symmetry that anticommutes with the mass term, satisfying
\begin{equation}
\{U_R,\sigma_x\}=0, \hspace{3mm} U_R \sigma_y=\eta \sigma_y U_R, 
\end{equation}
to meet the condition of Eq. (2). Moreover, spatial symmetries should be unitary and satisfy: 
\begin{equation}
[U_R, \mathcal{T}]=0, \hspace{3mm} U_R \mathcal{C}=\zeta \mathcal{C} U_R,
\end{equation}
with $\zeta=\pm 1$ determining whether the spatial symmetry commutes or anticommutes with chiral symmetry. Eqs. (3) and (4) necessitate $\zeta = -\eta$ for finding an appropriate matrix representation of spatial symmetries, specified by:
\begin{equation}
U_R=\pm [(\sigma_z + i \sigma_y)+\zeta (\sigma_z - i \sigma_y)]/2.
\end{equation}
This implies that a proper rotation should anticommute with the chiral symmetry, while an improper rotation should commute with it, enabling the emergence of intrinsic corner states in 2D BDI systems. 

\begin{figure*}[!t]
	\centering
	\includegraphics[scale=0.06]{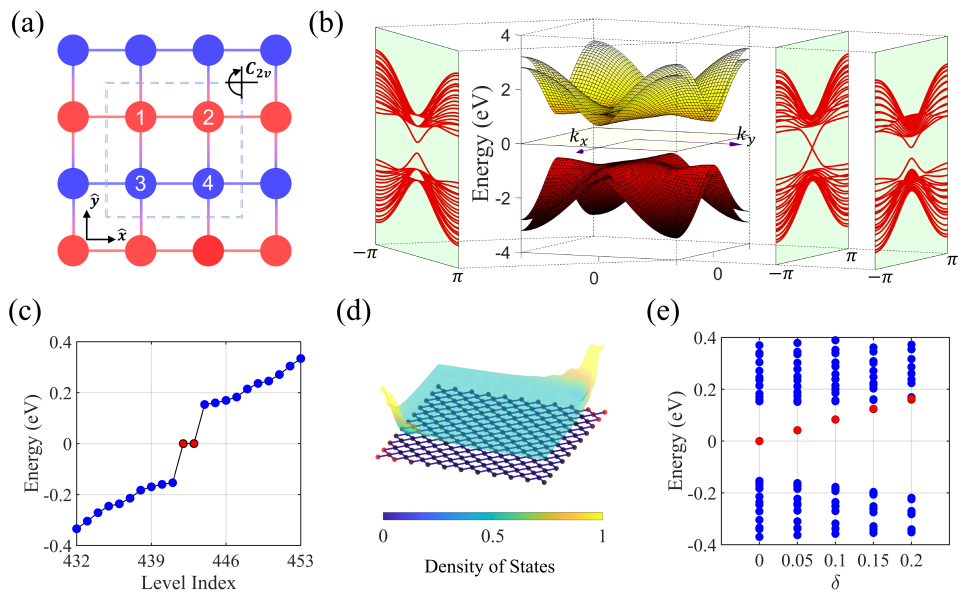}
    \caption{(a) Schematic of a two-dimensional lattice under point group $C_{2v}$ showing primitive cells (dashed square) and atoms categorized into distinct sublattice groups (red and blue circles). (b) Three-dimensional representation of bulk band structures from the tight-binding model (parameters $M=1$ and $m=0.2$). The left plot illustrates the edge states along the $xy$-direction, while the right plot depicts the edge states along the $x$-direction. The rightmost plot corresponds to $\delta=0.2$. (c) Energy levels of the tight-binding model for a nanodisk with 221 unit cells. The red points indicate the corner states. (d) Illustration of the nanodisk and the electronic charge density associated with the corner states. (e) Energy levels of the nanodisk as a function of $\delta$.}
\label{fig-2}
\end{figure*}

This analysis extends to 2D BDI systems under point group symmetries, where the potential positions of corner states are inferred from the relationships between the generators and chiral symmetry, as summarized in Fig. \ref{fig-1} (c). In two dimensions, improper rotations refer to mirror symmetries, while proper rotations relate to $C_n$ symmetries for $n=2,3,4,6$. Along the edges of these systems, specific positions may be invariant under mirror reflections but not under rotations. Consequently, in point groups governed solely by rotation symmetries, the precise positions of intrinsic corner states remain unpredictable, although the total number of these states is conserved due to the bulk topological phase. Notably, Eq. (5) does not account for order-three operations, which require $U_R^3 = \pm I$, thereby indicating that $C_3$ rotations are incapable of generating intrinsic corner states. In point groups with mirror symmetries, intrinsic corner states are located at the mirror-invariant corners, where the mirror symmetries commute with the chiral symmetry. For point groups featuring both mirror and rotational symmetries, the generators can be chosen as the minimal rotation and a mirror, leading to scenarios where either all or only half of the mirror symmetries commute with the chiral symmetry, resulting in two distinct potential configurations for intrinsic corner states.

To illustrate these symmetry constraints, we consider a 2D lattice with point group $C_{2v}$, as depicted in Fig. \ref{fig-2} (a). In crystals, chiral symmetry is typically achieved through sublattice arrangements with interactions confined to distinct sublattice groups, thus also being referred to as sublattice symmetry \cite{chiu2016classification}. The
combination of sublattice and time-reversal symmetries forms the particle-hole symmetry. The primitive cell contains four identical atoms, distributed into two sublattice groups, with the $s$ orbitals of each atom serving as basis functions. Spatial symmetry operators are defined as follows:
\begin{equation}
M_x = \tau_0\sigma_x, \hspace{3mm} M_y = \tau_x\sigma_0, \hspace{3mm} C_{2z} = \tau_x\sigma_x,
\end{equation}
where $\tau(\sigma)_i (i=x, y, z)$ are Pauli matrices. The nonspatial symmetries are represented by $\mathcal{T}=K$ in spinless systems and $\mathcal{S} = \tau_z\sigma_0$ with eigenvalues $\pm 1$ for basis functions in distinct sublattice groups. This setup ensures that spatial symmetries mapping atoms within the same sublattice group commute with the chiral symmetry, while the opposite mappings indicate anticommutation. Moreover, such a representation ensures that the Hamiltonian is block-anti-diagonalized. The Hamiltonian under these constraints satisfies the following relations: 
\begin{equation}
\begin{split}
\mathcal{T}H(\bm{k})\mathcal{T}^{-1}=H(-\bm{k})&, \hspace{3mm} \mathcal{S}H(\bm{k})\mathcal{S}^{-1}=-H(\bm{k}), \\
U_R H(\bm{k})U_R^{-1}&=H(R\bm{k}), \\
\end{split}  
\end{equation}
with $R$ representing a spatial symmetry and $U_R$ its matrix representation. The minimal Hamiltonian is expressed as:
\begin{equation}
H(k_x, k_y) = M\tau_x\sigma_x + k_x\tau_x\sigma_y + k_y\tau_y\sigma_0 + m\tau_x\sigma_0,
\end{equation}
where $M$ and $m$ are mass parameters in the bulk and along the edge, respectively. For any given edge orientation $\theta$, the momentum components parallel ($k_\parallel$) and perpendicular ($k_{\perp}$) to the edge are considered. The edge Hamiltonian then reads:
\begin{equation}
H_{\text{Edge}}(\theta, k_{\parallel}) = m \cos(\theta) \sigma_x + k_{\parallel} \sigma_y,
\end{equation}
aligning with Eq. (1) [refer to the Supporting Information (SI) for a detailed derivation]. Notably, symmetry-preserved terms in the bulk allow for a continuous transformation of the edge Hamiltonian, making it sufficient to consider only its minimal form. The mass term creates domain walls at $\theta$ values of $\frac{\pi}{2}$ and $\frac{3 \pi}{2}$, corresponding to the $M_x$-invariant points along the edge, as indicated in the $C_{2v}^-$ case in Fig. \ref{fig-1} (b), where $M_x$ commutes with the chiral symmetry, while $C_{2z}$ and $M_y$ anticommute with it.

To simulate the situation in a periodic lattice, we incorporate periodicity into the Hamiltonian from Eq. (7), reformulated within a tight-binding model as follows:
\begin{equation}
\begin{split}
H(k_x, k_y) = [M&-(2-\cos k_x-\cos k_y)]\tau_x\sigma_x \\
+& \sin k_x\tau_x\sigma_y + \sin k_y\tau_y\sigma_0 + m\tau_x\sigma_0.
\end{split}
\end{equation}
As shown in Fig. \ref{fig-2} (b), this model predicts a gapped bulk band structure across the entire Brillouin zone. Edge calculations reveal that edge states emerge within the gap of the projected bulk states, which are gapped along the $xy$-direction and gapless along the $x$-direction. This gapless nature is safeguarded by the $M_x$ symmetry, which commutes with the chiral symmetry, implying the presence of intrinsic corner states at the $M_x$-invariant corners. In Fig. \ref{fig-2} (c), energy level calculations for a nanodisk comprising 221 unit cells identify two corner states exactly at zero energy, with their electronic charge densities highly localized at two $M_x$-invariant corners in Fig. \ref{fig-2} (d).

Given that the chiral operator anticommutes with the Hamiltonian, each eigenstate is paired with a counterpart of opposite eigenvalue, resulting in the energy spectra exhibiting vertical symmetry about zero energy, as shown in Fig. \ref{fig-2} (b) and (c). This symmetry serves as a criterion for determining the presence of chiral symmetry in the system. When chiral symmetry is broken, such as through hoppings within the same sublattice, this symmetric pattern is disrupted. We introduce a $\delta \tau_x \sigma_0$ term to induce this chiral symmetry breaking. As shown in Fig. \ref{fig-2} (b), when $\delta=0.2$, the edge states are no longer symmetric about zero energy, and a gap opens in the spectrum. The energy levels in Fig. \ref{fig-2} (e) show that the corner states move away from zero energy as $\delta$ increases, eventually merging into the bulk states. In realistic materials, such symmetry-disrupting terms are hard to eliminate, but their influence can be minimal if $\delta \to 0$, which is a condition referred to as the ``sublattice approximation''.

\begin{figure}[!t]
	\centering
	\includegraphics[scale=0.056]{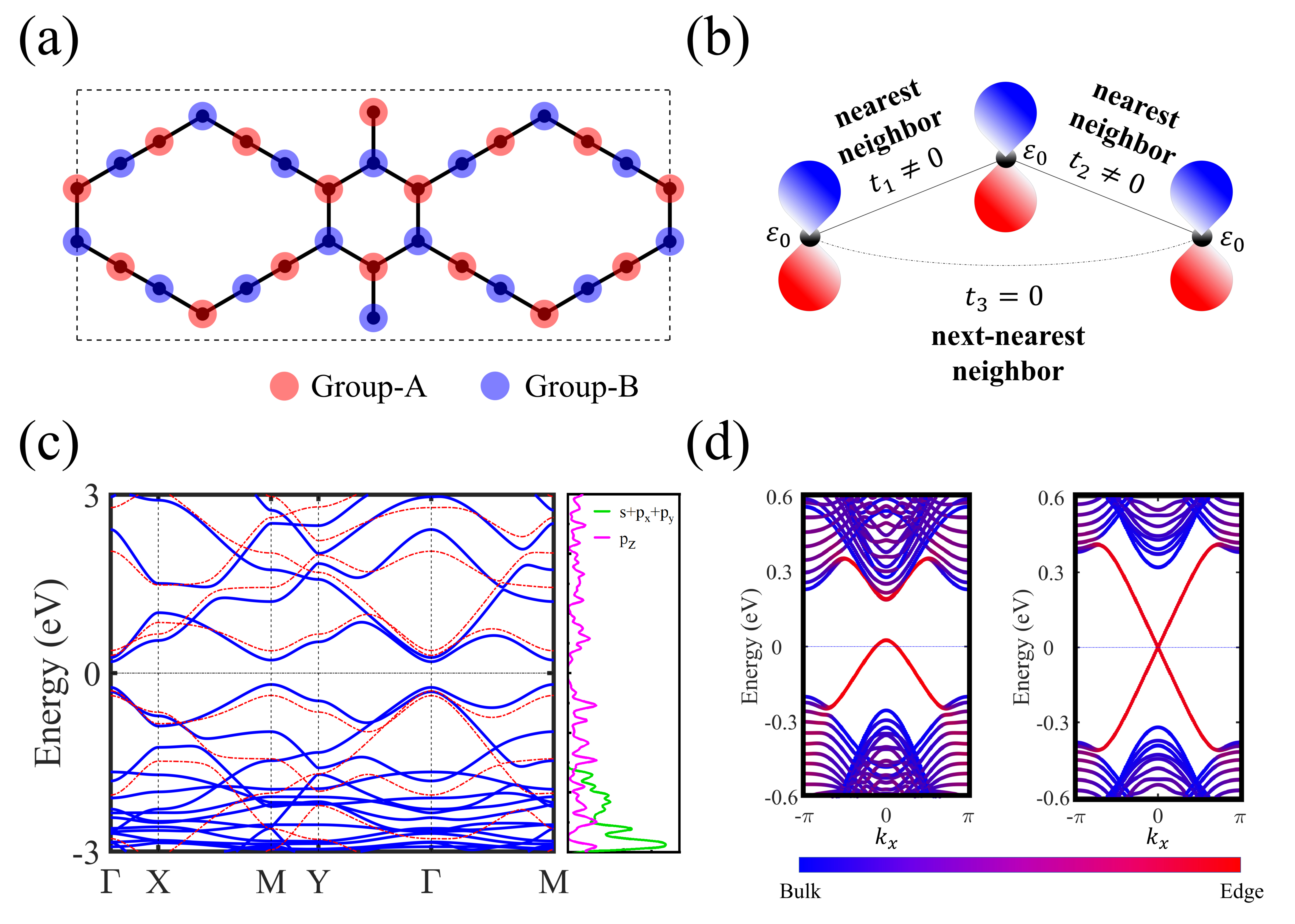}
    \caption{(a) Lattice structure of GY-1-5. The dashed box represents the primitive cell. The Wyckoff positions of the atoms include $2h$ at (0.5,0.29) and (0.5,0.91), $2g$ at (0.0,0.39), and $4i$ at (0.42,0.40), (0.35,0.71), (0.29,0.79), (0.21,0.90), (0.14,0.21), and (0.07,0.29). Red and blue circles represent atoms in distinct sublattice groups. (b) Schematic diagram of onsite energies ($\varepsilon_0$) and hopping parameters ($t_1$, $t_2$ and $t_3$) between the nearest and next-nearest neighbor $p_z$ orbitals with sublattice approximation. (c) Band structures of GY-1-5 without (solid blue lines) and with (red dashed lines) sublattice approximation. The right panel illustrates the partial density of states for the $p_z$ and other orbitals of GY-1-5. (d) Edge states along the $x$-direction of GY-1-5 without (left panel) and with (right panel) sublattice approximation. }
\label{fig-3}
\end{figure}

While particle-hole and chiral symmetries can be strictly enforced in simplified tight-binding models, reproducing these conditions in actual materials poses significant challenges. To demonstrate the effect of these symmetries in crystals, we employ a sublattice approximation, exemplified through our investigation of carbon-based polymorphic graphyne, known for its $sp$ and $sp_2$ hybridized bonds and considerable material design potential. Our prior work has extensively detailed the structural construction methodology and identified several structurally stable materials \cite{xu2023toward}. Herein, we delve into the higher-order topological properties of GY-1-5, a representative polymorphic graphyne configuration, 
to underscore its utility in studying intrinsic 2D SOTIs.
Additional materials can be found in the SI. 

The lattice of GY-1-5 belongs to the plane group \textsl{p2mm} and contains  30 atoms in its primitive cell, as shown in Fig. \ref{fig-3} (a).  The optimized lattice constants are $a=16.38$ \AA and $b=6.91$ \AA. The band structure of GY-1-5 is illustrated in Fig. \ref{fig-3} (c), which exhibits a semiconductor band gap of 0.38 eV. Near the Fermi level, we observe a nearly vertically symmetric pattern, indicating the presence of approximate sublattice symmetry in this material. In Fig. \ref{fig-3} (c), we present the projected density of states (PDOS) for the $s$ and $p$ orbitals, emphasizing that the bands near the Fermi level primarily originate from the $p_z$ orbital. Due to their high localization, non-nearest neighbor hoppings 
are substantially weaker than those between nearest neighbors. Given the uniformity of the element constituting the atoms, we can approximate that all $p_z$ orbitals possess similar onsite energies, allowing us to categorize the atoms into two distinct sublattice groups, as marked in Fig. \ref{fig-3} (a), thus ensuring minimal interactions within the same group.

\begin{figure}[!t]
	\centering
	\includegraphics[scale=0.045]{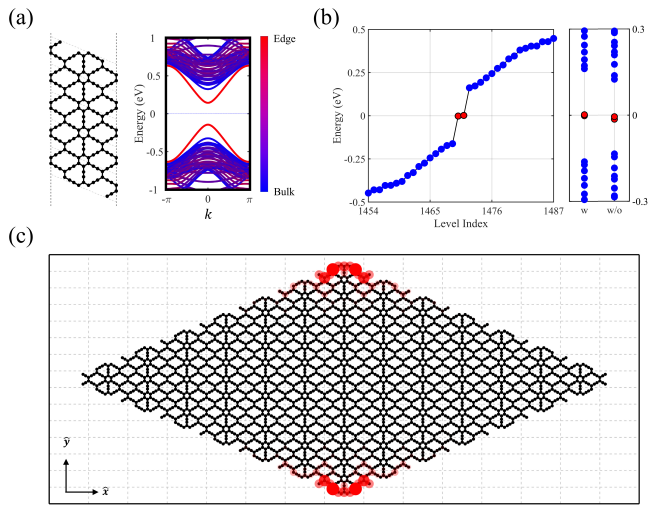}
    \caption{(a) Edge states of GY-1-5 along the $xy$-direction with sublattice approximation. The cutting edge is shown in the left panel. (b) Energy levels of a nanodisk with 2940 atoms with sublattice approximation. The red points indicate the intrinsic corner states. The right panel compares the energy levels with (w) and without (w/o) sublattice approximation. (c) Illustration of the nanodisk and the density of states for the intrinsic corner states.}
\label{fig-4}
\end{figure}

To compare the cases without and with sublattice approximation, we construct a Wannier tight-binding (TB) Hamiltonian using the maximally localized Wannier functions (MLWF) method with the WANNIER90 package \cite{mostofi2008wannier90,mostofi2014updated}. The band structure of GY-1-5 with sublattice approximation is shown by the dashed lines in  Fig. \ref{fig-3} (c), displaying an accurate vertically symmetric pattern. This approximation is introduced by retaining only the energy bands contributed by $p_z$ orbitals, preserving solely nearest-neighbor interactions, and unifying all on-site energies, as depicted in Fig. \ref{fig-3} (b). As previously analyzed, intrinsic corner states protected by the point group \textsl{2mm} should appear at the mirror-invariant corners, where the mirror operation commutes with the chiral symmetry. As shown in Fig. \ref{fig-3} (a), the effective mirror operation is $M_x$, as it maps atoms within the same sublattice group. As shown in Fig. \ref{fig-3} (d), the edge states along the $x$-direction transition from gapped to gapless when the sublattice approximation is applied, implying the presence of intrinsic corner states at the $M_x$-invariant corners.

To intuitively display the intrinsic corner states, we construct a nanodisk with $M_x$-invariant corners aligned with the edges in the $xy$-direction. Using the TB Hamiltonian with sublattice approximation, we show the gapped edge states along the $xy$-direction in  Fig. \ref{fig-4} (a), which corroborates the edge Hamiltonian described by Eq. (1). We then calculate the energy spectrum of the nanodisk, as shown in Fig. \ref{fig-4} (b), where two states emerge precisely at the Fermi level, representing the zero modes of the intrinsic 2D SOTIs. These states are firmly anchored at the Fermi level due to chiral symmetry constraints, with minor energy variations attributable to finite-size effects that diminish as the nanodisk size increases. The electronic charge density of these zero modes is shown in Fig. \ref{fig-4} (c), which exhibits pronounced localization at the two $M_x$-invariant corners. In the realistic case without sublattice approximation, the energy levels of corner states deviate slightly from the Fermi level, as shown in Fig. \ref{fig-4} (b). Nevertheless, the corner states remain well within the central region of the bandgap, which is a consequence of the approximate chiral symmetry in this material, thereby facilitating observation.

In conclusion, our work provides valuable insights into the underlying principles for  realizing intrinsic second-order topological insulators (SOTIs) in two dimensions, particularly within class BDI. We establish the symmetry conditions essential for the emergence of intrinsic corner states and offer practical guidelines for the material realization of intrinsic SOTIs using sublattice approximation. Our theoretical arguments are supported by first-principles calculations on polymorphic graphyne materials, which exhibit the potential to generate diverse symmetries through artificial design. These materials are ideal candidates for investgating the topological properties of intrinsic 2D SOTIs and for facilitating the observation of topological corner states. 

This work is supported by the National Natural Science Foundation of China (Grant No. 12188101,  12204224, 12247138, 11925408 and 11921004, ), the Guangdong Basic and Applied Basic Research Foundation (Grant No. 2023A1515010734), the Ministry of Science and Technology of China (Grant No. 2022YFA1403800), the Chinese Academy of Sciences (Grant No. XDB33000000), the New Cornerstone Science Foundation through the XPLORER PRIZE and the Major Science and Technology Infrastructure Project of Material Genome Big-science Facilities Platform supported by Municipal Development and Reform Commission of Shenzhen.

Z. J. Chen and S. G. Xu contributed equally to this work.


\end{document}